# The energy representation of world GDP

Boris M. Dolgonosov

Haifa, Israel

borismd31@gmail.com

**Abstract.** The dependence of world GDP on current energy consumption and total energy produced over the previous period and materialized in the form of production infrastructure is studied. The dependence describes empirical data with high accuracy over the entire observation interval 1965 – 2018.

World economy is driven by energy which is used for creating production infrastructure and for its operation. The existing global infrastructure has an energy equivalent equal to the total energy consumed in all previous years. In other words, previously used energy is materialized in the created infrastructure, whose operation is ensured by current energy consumption. This gives reason to look at the production function from an energy point of view and establish parallels between energy and such classical concepts as labor and capital, which, following Cobb and Douglas (1928), determine the production function. More detailed models expanding the Solow's (1956) theory consider labor, capital, and energy as independent quantities (Stern and Kander 2011, Csereklyei et al. 2016, Gozgor et al. 2018), although this complicates the description due to the abundance of parameters. Energy representation, introduced in (Dolgonosov 2018), gives a simpler description, at least for the global scale.

The objective of this communication is to analyze the energy representation of world GDP and verify its compliance with empirical data. The global approach allows us to exclude from consideration energy flows between parts of the system.

The energy representation (Dolgonosov 2018) has the form

$$G = g(EU)^\gamma, \qquad (1)$$

where $G$ is world GDP, $g$ and $\gamma$ are constants, $E$ is total energy consumption, and $U$ is materialized energy, defined as energy consumed for all previous years and used to create production infrastructure:

$$U(t) = \int_{-\infty}^{t} E(x)dx. \tag{2}$$

It is assumed that energy consumption rapidly decreases with tending to the past, ensuring the existence of the improper integral (2).

Representation (1) is similar to a Cobb and Douglas (1928) production function in the form

$$G = AL^{\alpha}K^{\beta}, \tag{3}$$

where $A$ is a coefficient, $L$ = labor, $K$ = capital, $\alpha, \beta \in [0,1]$. Relation (1) is a special case of equation (3), when $\alpha = \beta = \gamma$, $L \propto E$ (labor costs are proportional to energy consumption) and $K \propto U$ (capital is proportional to materialized energy). Indeed, labor uses energy $E$ to operate the infrastructure, while capital is embodied in the infrastructure itself, which was created using the total energy $U$ produced over all previous years.

We can also arrive at relation (1) in another way. Consider the case when GDP growth occurs over a certain characteristic time $\tau$ (for example, if the process develops exponentially). Then from the dimensional quantities $E, U, \tau$ that characterize the production system, we can compose the only combination in units of energy consumption (power): $(EU/\tau)^{1/2}$. Multiplying this combination by a coefficient $g_1$, which converts energy units into financial ones, we obtain $G = g_1(EU/\tau)^{1/2}$ that corresponds to equation (1) with the exponent $\gamma = 1/2$.

The situation changes if GDP growth does not have a characteristic time (for example, if the process develops according to a power law). This possibility follows from the relationship $G = pN^q$ (Fig. 1f) and the hyperbolic law $N = C/(t_s - t)$, where $N$ is world population, $q \approx 2$, $p$ and $C$ are constants, $t_s$ is the singularity moment, which according to Foerster et al. (1960) falls at the end of 2026. Accordingly, GDP obeys the law $G \propto (t_s - t)^{-q}$. It is clear that the interval $t_s - t$ cannot represent the characteristic time of the process, since it is not constant. The singularity is the result of idealization; in fact, hyperbolic growth should cease as it approaches $t_s$. Demographic data show that the hyperbolic law has acted for a long time (about a thousand years), and is currently being violated, showing a decrease in growth rate (Akaev and Sadovnichii 2010, Korotayev et al. 2015, Dolgonosov 2016). In the absence of characteristic time, the process is significantly extended, amplifying the effect of energy on GDP that leads to a higher value of the exponent in equation (1): $\gamma > 1/2$.

To find the parameters of model (1), we used the literature data on world GDP (World Bank 2020), energy consumption (BP 2020), and population (US Census Bureau 2020), which are shown in Fig. 1a. We use the following units: time $t$, year; energy consumption $E$, Gtoe/year; materialized energy $U$, Gtoe; GDP, T\$/year (where Gtoe = Giga ton of oil equivalent, T\$ = Trillion \$ US 2010). The choice of the time interval 1965 – 2018 (54 years) is dictated by the availability of systematic data on energy consumption. In calculations, integral (2) is replaced by summation over the years:

$$U_t = U_0 + \sum_{i=i_0}^{t} E_i \qquad (4)$$

where $i_0 = 1965$ is the initial year, $U_0$ is energy materialized in previous years (till year $i_0$). Calibration of model (1) and (4) shows that the minimum standard deviation of $G$ from GDP data (equal to 0.52 Gtoe/year or 1.2% of the average) is achieved at

$$U_0 = 141.25 \text{ Gtoe}. \qquad (5)$$

According to equation (1), the relationship between the logarithms of $G$ and $EU$ should be linear, which is confirmed with high accuracy: linear regression fits ideally on empirical data (Fig. 1b). Based on the indicated regression, one can find the parameter values in (1):

$$g = e^{-1.2459} = 0.2877, \ \gamma = 0.6258; \qquad (6)$$

$g$ is a dimensional quantity expressed here in the system of units Gtoe, T\$, year.

The GDP calculated by equations (1), (4) – (6) is in good agreement with empirical data (Fig. 1c). The invariability of $g$ and $\gamma$ throughout the entire time interval suggests that there is a time-independent relationship between energy and GDP, although the basic quantities vary significantly during this time (Table 1). It is impossible to use the ratio $E$/GDP (= energy intensity) for this purpose, due to the fact that energy intensity is not constant, but rather decreases significantly over time (1.5 times for the period under consideration) (Table 1; see also: Csereklyei et al. 2016, Stern 2018).

Equation (1) can also be obtained by considering $E$, $U$, and GDP as functions of the population. These dependencies are shown in Fig. 1d, e, f. As seen, the following power-law approximations have good accuracy:

$$E = kN^d, k = 0.7121, d = 1.4596, \tag{7}$$

$$U = rN^s, r = 16.331, s = 1.7632, \tag{8}$$

$$G = pN^q, p = 1.3344, q = 2.0175. \tag{9}$$

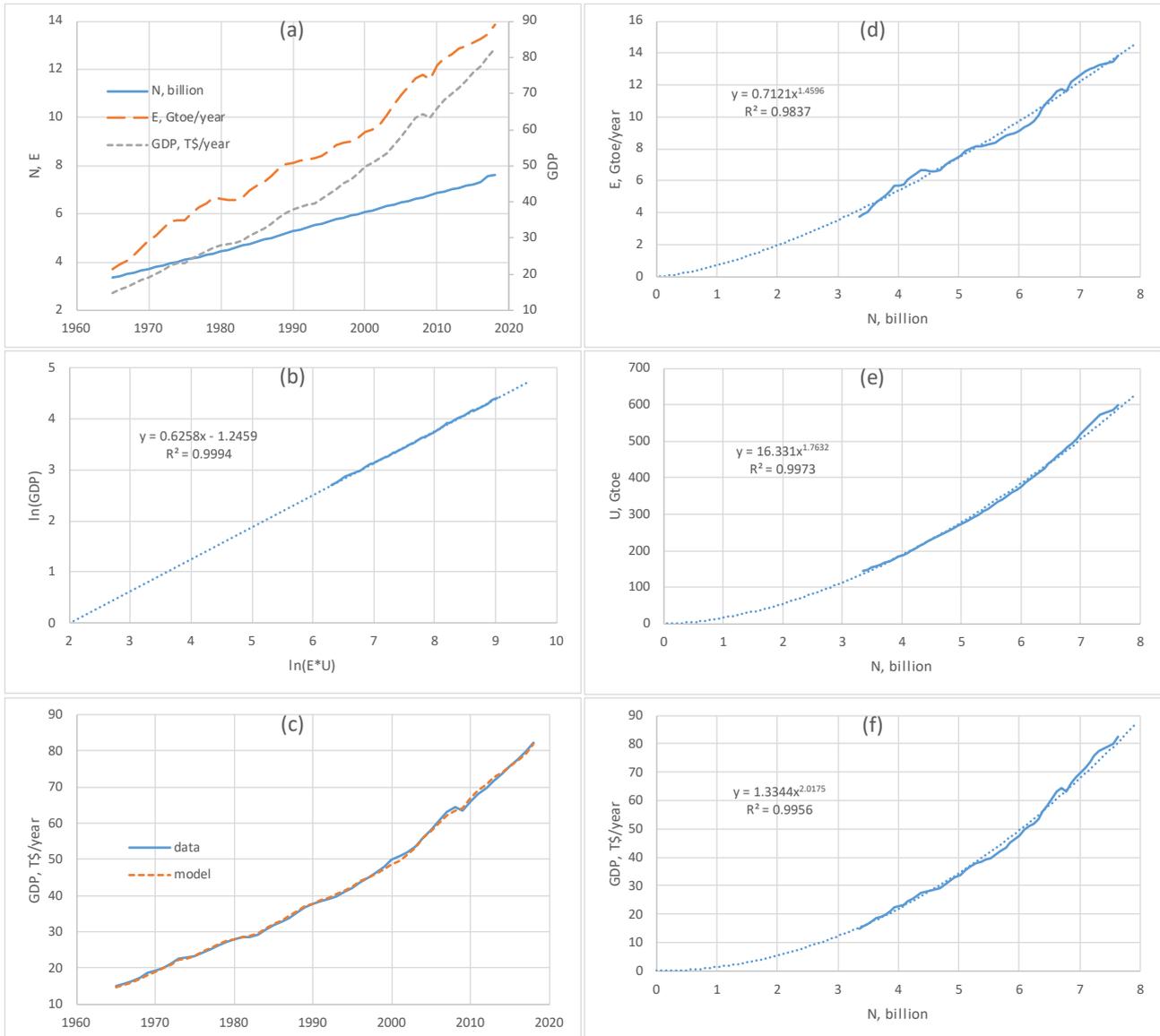

Fig. 1.

(a) World population, energy consumption, and GDP over 1965 – 2018. Data sources: $N$ – (US Census Bureau 2020), $E$ – (BP 2020), GDP – (World Bank 2020).

(b) log (GDP) versus log ($EU$).

(c) Comparison of calculated GDP with data.

(d, e, f) Energy consumption ($E$), materialized energy ($U$), and GDP as functions of world population ($N$).

Table 1. Changes in basic quantities

| Year | E | GDP | E/GDP | U |
|---|---|---|---|---|
| 1965 | 3.73 | 14.84 | 0.251 | 145.0 |
| 2018 | 13.86 | 82.46 | 0.168 | 600.7 |
| 2018 to 1965 | 3.72 | 5.56 | 0.669 | 4.14 |

Substituting (7) – (9) into (1), we obtain relationships between the parameters

$$p = g(kr)^\gamma, \quad q = \gamma(d + s) \tag{10}$$

and find the values $p = 1.3358$ and $q = 2.0168$, which are very close to their values (9), obtained independently through the regression in Fig. 1f. This is additional evidence in favor of key equation (1).

Thus, the energy representation of GDP (1) receives versatile confirmations.

**Conclusions**

1. World GDP is completely determined by two energy quantities: current energy consumption $E$ and total energy $U$ produced in the previous time and materialized in the form of production infrastructure.

2. GDP is a power-law function of the combined variable which is the product of energy consumption and materialized energy, $G = g(EU)^\gamma$.

3. The materialized energy $U_t$ is calculated as the sum of the initial value $U_0$ and the subsequent accumulation $\sum_i E_i$ up to year $t$.

4. Calibration of model (1) and (4) according to empirical data over 1965 – 2018 gives the parameter values: $U_0 = 141.25$ Gtoe (to top 1965), $\gamma = 0.6258$, $g = 0.2877$ (the latter quantity is dimensional, expressed in the system of units: Gtoe, T$ US 2010, year).

5. Model (1) and (4) describes empirical data with high accuracy over the entire observation interval 1965 – 2018, during which energy and GDP change significantly.

6. Energy consumption, materialized energy, and GDP are power-law functions of the population. Parameters of these functions are interconnected by two relations (10), which follow from model equation (1).